\documentclass[doublecol,linenumbers]{epl2} 

\usepackage{amsmath}
\usepackage{amssymb}
\usepackage{graphicx}
\usepackage{dcolumn}
\usepackage{bm}
\usepackage{setspace}
\usepackage{amsfonts}
\usepackage{rotating}
\usepackage{makeidx}
\usepackage{amsthm}

\newcommand{\openone}{\mathbb{I}}

\title{Assisted optimal state discrimination without entanglement \footnote{EPL, 106 (2014) 50004.  doi: 10.1209/0295-5075/106/50004}}
\shorttitle{Assisted optimal state discrimination without entanglement} 

\author{L. F. Xu\inst{1} \and F. L. Zhang\inst{1}\footnote{Corresponding author: flzhang@tju.edu.cn} \and M. L. Liang\inst{1} \and J. L. Chen\inst{2,3}}
\shortauthor{L.F. Xu \etal}

\institute{
  \inst{1} Department of Physics, School of Science, Tianjin University, Tianjin 300072, China\\
  \inst{2} Theoretical Physics Division, Chern Institute of Mathematics, Nankai University, Tianjin, 300071, China\\
  \inst{3} Centre for Quantum Technologies, National University of Singapore, 3 Science Drive 2, Singapore 117543
}


\abstract{
A fundamental problem in quantum information is to explore the roles of different quantum correlations in a quantum information procedure.
Recent work [Phys. Rev. Lett., 107 (2011) 080401] shows that the protocol for assisted optimal state discrimination (AOSD) may be implemented successfully without entanglement, but with another correlation, quantum dissonance.
However, both the original work and the extension to discrimination of $d$ states [Phys. Rev. A, 85 (2012) 022328] have only proved that entanglement can be absent in the case with equal a \emph{priori} probabilities.
By improving the protocol in [Sci. Rep., 3 (2013) 2134], we investigate this topic in a simple case to discriminate three nonorthogonal states of a qutrit, with positive real overlaps.
In our procedure, the entanglement between the qutrit and an auxiliary qubit is found to be completely unnecessary.
This result shows that the quantum dissonance may play as a key role in optimal state discrimination assisted by a qubit for more general cases.
}

\begin{document}

\maketitle

\section{Introduction}

  Quantum correlations contained in composite quantum states play important roles in quantum information processing and have been widely studied from various perspectives.
Many concepts have been presented to reflect these correlations, such as quantum entanglement \cite{Horodecki09}, Bell nonlocality \cite{Bell}, and quantum discord \cite{PhysRevLett.88.017901,henderson2001classical}.
Entanglement had been regarded as the only resource for demonstrating the superiority of quantum information processing \cite{Horodecki09,key}.
However, recent studies \cite{lanyon2008experimental,datta2008quantum} show that the
algorithm for deterministic quantum computation with one qubit (DQC1) can surpass the performance of the corresponding classical algorithm in the absence of entanglement between the control qubit and a completely mixed state.
The quantum discord, which measures the nonclassical correlations and can exist in a separable state, is regarded to be the key resource in this quantum algorithm and has gained wide attention in recent years.
Based on a unified view \cite{modi2010unified} of quantum and classical correlations, another type of quantum correlations called dissonance was put forward.
Quantum dissonance measures the nonclassical correlations with entanglement being completely excluded.
For a separable state, its dissonance is exactly equal to the quantum discord.
Therefore, the quantum discord playing a key role in the computational process is nothing but the dissonance.

Recently, Roa \emph{et. al.} \cite{roa2011dissonance} provided another example after DQC1 in which quantum dissonance serves as the key resource.
Namely, they show that for performing assisted optimal state discrimination (AOSD), dissonance is the only quantum correlation required when two nonorthogonal states are prepared with equal \emph{a priori} probabilities.
This result has been extended to the case with $d$ nonorthogonal states \cite{BoLi}.
Another line of development in this topic has been towards extending this result to the cases with arbitrary \emph{priori} probabilities.
To this end, Zhang \emph{et. al.} \cite{Zhang} improved the protocol and proved the entanglement is completely unnecessary for the AOSD of two states.
That is, it is quantum dissonance that plays a key role in AOSD rather than quantum entanglement.

The above development inspires us to study this topic in high dimensions with arbitrary \emph{priori} probabilities.
However, it is not easy to obtain analytical solutions for the AOSD of more than two pure states\cite{PhysRevA.82.032338}.
In this work, we confine ourselves to the case of three linearly independent nonorthogonal states with positive real overlaps,
 which has been studied in the positive-operator-valued measure (POVM) formalism recently \cite{PhysRevA.82.032338}.
To achieve the optimal discrimination, we need to extend the protocol given by Zhang \emph{et. al.} \cite{Zhang} to a more general form [see Eq. (\ref{Utrans}) below]. For the optimal case, we found entanglement to be completely unnecessary and derived the quantum dissonance by using the geometric measure of quantum discord (GMQD) \cite{GMQD,GMQD2013}.
The details are shown in the following parts.

\section{AOSD of a qutrit}
Let us consider a three-dimensional system (principal qutrit) randomly prepared in one of the three nonorthogonal states $|\psi_i\rangle$,
with a \emph{priori} probabilities $p_i$, where $p_i\in [0,1]$ and $\sum_{i=0}^2 p_i=1$.
For simplicity, we set the inner products $\langle\psi_i|\psi_{j \neq i}\rangle=\gamma_{ij}  \in [0,1]$.
To discriminate these three states $\{|\psi_i\rangle\}$ unambiguously,  we couple the qutrit to an auxiliary qubit $A$, prepared in a known pure state $|k\rangle_a$.
Performing a joint unitary transformation $\mathcal {U}$ on their whole states, one obtains
\begin{eqnarray}\label{Utrans}
\mathcal{U}\;|\psi_i\rangle|k\rangle_a=\sqrt{1-|\alpha_i|^2}\;|i\rangle|0\rangle_a+\alpha_i|\Phi_i\rangle|1\rangle_a,
\end{eqnarray}
where $i=0,1,2$, and $\{|i\rangle\}$ and $\{|0\rangle_a,|1\rangle_a\}$ are the basis for the principal system and the ancilla respectively.
After the joint transformation, the mixed state we consider in discrimination is given by
\begin{eqnarray}\label{rho}
\rho_{SA}   =  \sum_{i=0}^2 p_i \;\mathcal {U}\left( |\psi_i\rangle\langle\psi_i|\otimes |k\rangle_a \langle k|\right) \mathcal {U}^\dag
\end{eqnarray}
The auxiliary system will collapses to either $|0\rangle_a $ or $|1\rangle_a$
by performing a von Neumann measurement on the basis $\{|0\rangle_a , |1\rangle_a\}$.
If the auxiliary system collapses to $|0\rangle_a $, the state discrimination is successful, because the states $|i\rangle$ in (\ref{Utrans}) can be distinguished deterministically.
Otherwise, we fail when the qubit collapses to $|1\rangle_a$.
The distinguishing probability of success is then
\begin{eqnarray}\label{prob}
P_{\rm suc}&=&{\rm Tr}[(\openone_s\otimes|0\rangle_a\langle0|)\rho_{SA}]\nonumber\\
&=&1-\sum_{i=0}^2 p_i|\alpha_i|^2 = 1- \sum_{i=0}^2 p_i\gamma_i\biggr|\frac{J_{jk}}{{J_{ij}J_{k i}}}\biggr|  ,
\end{eqnarray}
where $(i,j,k)=(0,1,2),(1,2,0)$ or $(2,0,1)$,  $\gamma_i= \gamma_{ij}\gamma_{k i} / \gamma_{jk}  $, and $\openone_s$ is the unit matrix for the principal qutrit.
Here, we set $J_{ij}=\langle\Phi_i|\Phi_j\rangle$, and they satisfy $\alpha_i^\ast\alpha_j J_{ij}=\gamma_{ij}$.

An improvement of (\ref{Utrans}) to the previous protocols \cite{roa2011dissonance,BoLi,Zhang} is that we allow the difference among states $|\Phi_i\rangle$.
Applying the general transformation to the case of two states\cite{roa2011dissonance,Zhang}, one can find the optimal discrimination when $|J_{01}|=1$.
Namely, for a fixed $J_{01}$, the maximum of $P_{\rm suc} = 1- p_0 |\alpha_0|^2 - p_1 |\alpha_1|^2$ under the constraint $|\alpha_0 \alpha_1 J_{01}|=\gamma_{01}$ can be derived as $P_{\rm suc}=1-2 \sqrt{p_0p_1} |\gamma_{01}/J_{01}|$ or $\max\{p_0,p_1\}(1-|\gamma_{01}/J_{01}|^2)$ by using the method of  \cite{roa2011dissonance,Zhang}. Obviously, when $|J_{01}|=1$, $P_{\rm suc}$ achieves its optimum.

For the case of three states with positive real overlaps, to derive the optimal process of state discrimination and study the roles of quantum correlations, we write
the states $|\Phi_i\rangle$ as
\begin{eqnarray}\label{Phi}
&&|\Phi_0\rangle =|\eta_0\rangle, \ \ \ \ \ \
|\Phi_1\rangle =\cos \theta_1 |\eta_0 \rangle+ \sin \theta_1 |\eta_1\rangle,   \\
&&
|\Phi_2\rangle =\cos \theta_2 |\eta_0 \rangle+ \sin \theta_2 \cos \theta_3 e^{i \phi}|\eta_1\rangle+ \sin \theta_2\sin \theta_3 |\eta_2\rangle, \nonumber
\end{eqnarray}
where $\{|\eta_i\rangle\}$ is a basis for the principal qutrit.
Obviously, the success probability  $P_{\rm suc}$ is independent of the forms of $ |\eta_i\rangle $.
Thus, we can first determine the maximum value of $P_{\rm suc}$ and the corresponding parameters $\alpha_{i}$, $\theta_{j}$, and $\phi$ in this part,
 and investigate the quantum correlations in state $\rho_{SA} $ by adjusting $|\eta_i\rangle $ in next section.
For simplicity and without loss of generality, we assume ${p_0 \gamma_0}\geq {p_1 \gamma_1}\geq {p_2 \gamma_2}$.
The problem to maximize the discrimination success probability can be solved in the following two steps.

\emph{Step 1:}
Among the region of $\theta_j,\phi \in [0,2\pi)$, the discrimination success probability $P_{\rm suc} =1- \sum_{i=0}^2 p_i\gamma_i |{J_{jk}}/{({J_{ij}J_{k i}})} |  $ can be maximized as
\begin{equation}
P_{\rm suc, max}^{(i)}=1- ( p_0 \gamma_0+p_1 \gamma_1+p_2 \gamma_2) \nonumber \\
\end{equation}
when $\sqrt{p_0\gamma_0} \leq \sqrt{p_1\gamma_1} + \sqrt{p_2\gamma_2}$, which was called the \emph{triangle condition} in \cite{PhysRevA.82.032338}, and
\begin{equation}
P_{\rm suc, max}^{(ii)}=1-2(\sqrt{p_0p_1\gamma_0\gamma_1}+\sqrt{p_2p_0\gamma_2\gamma_0}-\sqrt{p_1p_2\gamma_1\gamma_2}),\nonumber
\end{equation}
for$\sqrt{p_0\gamma_0} > \sqrt{p_1\gamma_1} + \sqrt{p_2\gamma_2}$.
The first case occurs at $ |\Phi_0\rangle=|\Phi_1\rangle=|\Phi_2\rangle $ with $\theta_1 = \theta_2 =0$.
But the nonuniformity of $ p_j \gamma_j$  leads to a symmetry breaking of $ \{ |\Phi_j\rangle \}$ in second one, where $\theta_1 + \theta_2 = f(p_0\gamma_0,p_1\gamma_1,\Upsilon), \theta_2 = f(p_1\gamma_1,p_0\gamma_0,\Upsilon), \theta_3=\pi$, and $\phi=0$.
 Here, $\Upsilon=\sqrt{p_2\gamma_2}/( \sqrt{p_0\gamma_0} - \sqrt{p_1\gamma_1})$ and $f(r,s,t)=\arccos \sqrt{1/[1+\sqrt{\frac{r}{s}}(\frac{1}{t}-1)]}$.
Let us denote the values of $ \sqrt{\gamma_i |{J_{jk}}/{({J_{ij}J_{k i}})} |} $ in the two cases as $\bar{\alpha}_i$.
When $ \bar{\alpha}_i  \leq1$, the optimal discrimination probabilities take the forms of  $P_{\rm suc, max}^{(i)}$ and $P_{\rm suc, max}^{(ii)}$ respectively, and $|\alpha_i|= \bar{\alpha}_i $. There are two of $\bar{\alpha}_i>1$ at most.
If one of $\bar{\alpha}_i>1$, the optimal discrimination probability can be derived in \emph{Step 2}.
When two of $\bar{\alpha}_i>1$, two candidates of the optimal success probability can be found in \emph{Step 2}, and the larger one is the solution of the problem.

\emph{Step 2:}
Assuming only $\bar{\alpha}_2>1$, the success probability can be optimized by omitting the corresponding state $|\psi_i \rangle$ in the  discrimination procedure.
Namely, one can maximize  the success probability $P_{\rm suc} =1- \sum_{i=0}^2 p_i\gamma_i |{J_{jk}}/{({J_{ij}J_{k i}})} |  $ within the region of $\theta_j,\phi \in [0,2\pi)$ and under the constraint $|\alpha_2|=\sqrt{\gamma_2 |{J_{01}}/{({J_{12}J_{20}})} |}=1$ as
\begin{equation}
P_{\rm suc, max}^{(iii)}= 1-p_2 -2\sqrt{p_0p_1\gamma_0 \gamma_1}   - (\sqrt{p_0 \gamma_0}-\sqrt{p_1 \gamma_1})^2\gamma_2, \nonumber
\end{equation}
when $\phi=0$,  $\theta_1 + \theta_2 = f(p_0\gamma_0,p_1\gamma_1,\gamma_2)$, $\theta_2 = f(p_1\gamma_1,p_0\gamma_0,\gamma_2)$, and $\theta_3=\pi$.
Then, if the corresponding values of $ \tilde{\alpha}_i= \sqrt{\gamma_i |{J_{jk}}/{({J_{ij}J_{k i}})} |}  \leq 1$ for $(i,j,k)=(0,1,2)$ and $(1,2,0)$, the  optimal discrimination probability takes the form of $P_{\rm suc, max}^{(iii)}$.
However, when one of $\tilde{\alpha}_{0,1}>1$,  we have to omit the corresponding state other than  $|\psi_2 \rangle$  in the protocol.
Let us assume $\tilde{\alpha}_{1}>1$. We derive the maximum of  $P_{\rm suc} =1- \sum_{i=0}^2 p_i\gamma_i |{J_{jk}}/{({J_{ij}J_{k i}})} |  $ within the region of $\theta_j,\phi \in [0,2\pi)$ and under the constraint
$|\alpha_1|=|\alpha_2|=1$ as
\begin{equation}
P_{\rm suc, max}^{(iv)}=1-p_1-p_2- p_0\gamma_0\frac{\gamma_1+\gamma_2-2\gamma_1\gamma_2}{1-\gamma_1\gamma_2}, \nonumber
\end{equation}
when $\cos \theta _{1}=\sqrt{{(1-\gamma _{1}\gamma _{2})\gamma _{1}}/{(\gamma _{1}+\gamma _{2}-2\gamma
_{1}\gamma _{2})}}$ and $\cos \theta _{2}=\sqrt{{(1-\gamma _{1}\gamma _{2})\gamma _{2}}/{(\gamma _{1}+\gamma _{2}-2\gamma
_{1}\gamma _{2})}}$.
Then the value of $|\alpha_0|^2=\gamma_0 {(\gamma_1+\gamma_2-2\gamma_1\gamma_2)}/{(1-\gamma_1\gamma_2)} \leq 1$.
If two of $\bar{\alpha}_i>1$, one can begin by omitting one of the two corresponding states respectively as the case with only $\bar{\alpha}_2>1$, and derive two candidates of the optimal success probability in the form as $P_{\rm suc, max}^{(iii)}$ or $P_{\rm suc, max}^{(iv)}$.


\begin{figure}
\includegraphics[width=8cm]{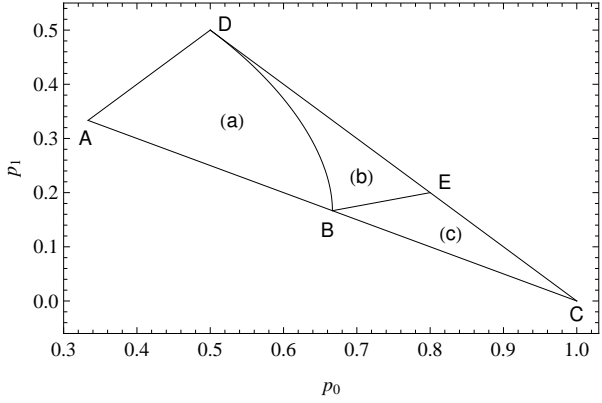} \\
 \caption{The region of a \emph{priori} probabilities under the constraint $p_0\geq p_1\geq p_2$ is plotted in the $(p_0, p_1)$-plane as the triangle $ACD$.
The points $A,C,D$ locate at $(1/3,1/3)$, $(1,0)$, and $(1/2,1/2)$. Two critical value of $\gamma$, $\gamma^c_{1}= \sqrt{p_2}/ (\sqrt{p_0}-\sqrt{p_1})$ and $\gamma^c_{2}= \sqrt{p_1}/(\sqrt{p_0}-\sqrt{p_1})$
divide the triangle $ACD$ into three parts as (a): $\gamma^c_{2} \geq \gamma^c_{1} \geq 1  $, (b):  $\gamma^c_{2} \geq 1 \geq \gamma^c_{1}  $, and (c): $1 \geq \gamma^c_{2} \geq \gamma^c_{1} $.
} \label{fig1}
\end{figure}

For the simple case with equal a \emph{priori} overlap $\gamma_{ij}=\gamma$, we assume ${p_0}\geq {p_1}\geq {p_2}$.
The region of the \emph{priori} probabilities can be plotted in the $(p_0, p_1)$-plane as the triangle $ACD$ shown in Fig. \ref{fig1}.
In the \emph{Step 1}, the triangle condition can be written as $\gamma^c_1= \sqrt{p_2}/ (\sqrt{p_0}-\sqrt{p_1})>1$, which divided the part (a) from the other ones in Fig. \ref{fig1}.
To derive the optimal discrimination, one has
\begin{equation}\label{Pm1}
P^{(i)}_{\rm suc, max}=1-\gamma,
\end{equation}
with $\bar{\alpha}^2_j= \gamma $, and
\begin{equation}\label{Pm2}
P^{(ii)}_{\rm suc, max}=1-2(\sqrt{p_0p_1}+\sqrt{p_0p_2}-\sqrt{p_1p_2})\gamma,
\end{equation}
with $\bar{\alpha}^2_0= \gamma(\sqrt{p_1}+\sqrt{p_2})/\sqrt{p_0}$, $\bar{\alpha}^2_1= \gamma(\sqrt{p_0}-\sqrt{p_2})/\sqrt{p_1}$, and $\bar{\alpha}^2_2= \gamma(\sqrt{p_0}-\sqrt{p_1})/\sqrt{p_2}$.
Obviously, for the first case the values of $\bar{\alpha}_j\leq 1$ for arbitrary $\gamma \in [0,1]$.
But, for the second, $\bar{\alpha}_j\leq 1$ require the overlap $\gamma \leq \gamma^c_1$.
Therefore, when $\gamma \leq \gamma^c_1\leq1$, the optimal discrimination success probability is $P^{(ii)}_{\rm suc, max}$ in (\ref{Pm2}),
and it takes the form as (\ref{Pm1}) if $\gamma^c_1>1$.

When $\gamma^c_1 < \gamma \leq 1$, one can obtain the optimal discrimination by using the \emph{Step 2} and omitting the state $|\psi_2\rangle$.
Then, we have
\begin{equation}\label{Pm3}
P^{(iii)}_{\rm suc, max}=1-p_2 -2\sqrt{p_0p_1}\gamma - (\sqrt{p_0}-\sqrt{p_1})^2\gamma^2,
\end{equation}
with $\tilde{\alpha}^2_0=\gamma [\sqrt{p_1}-(\sqrt{p_0}-\sqrt{p_1})\gamma]/\sqrt{p_0}$ and $\tilde{\alpha}^2_1=\gamma [\sqrt{p_1}-(\sqrt{p_0}-\sqrt{p_1})\gamma]/\sqrt{p_1}$.
One can easily find that, when $\gamma <\gamma^c_2 =\sqrt{p_1}/(\sqrt{p_0}-\sqrt{p_1})$, $\tilde{\alpha}_0 \leq \tilde{\alpha}_1 \leq 1$
and therefore $P^{(iii)}_{\rm suc, max}$ in (\ref{Pm3}) is the optimal success probability.
However, if $\gamma > \gamma^c_2$, $\tilde{\alpha}_1 > 1$, we have to omit the sates $|\psi_1\rangle$ and $|\psi_2\rangle$ simultaneously.
In this case, the optimal success probability is given by
\begin{equation}\label{Pm4}
P^{(iv)}_{\rm suc, max}=1-p_1-p_2- {2 p_0 \gamma^2}/{(\gamma+1)}.
\end{equation}
The region of the probabilities $\{p_j\}$ satisfying $\gamma^c_1\leq1$ in Fig. \ref{fig1} is divided into two parts by the critical value $\gamma^c_2$, which are (b) with $\gamma^c_2\geq1$ and (c) with $\gamma^c_2 < 1$.
Consequently, the form of the optimal probability (\ref{Pm4}) is absent in region (b).





\begin{figure}
\includegraphics[width=8cm]{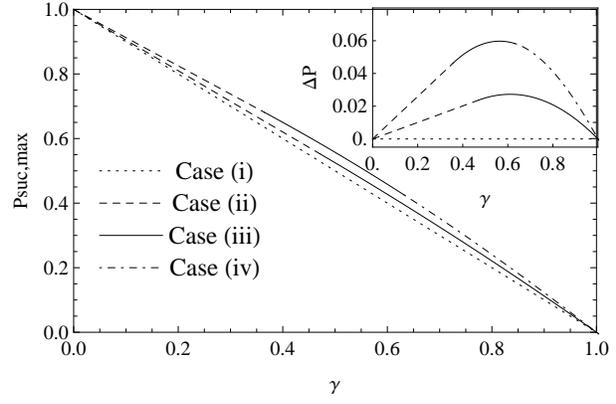} \\
 \caption{We plot the optimal success probability $P_{\rm suc, max}$ as a function of $\gamma$.
The \emph{priori} probabilities are chosen in the three regions in Fig. \ref{fig1}, in which the value of $(p_0,p_1)$ are : $(0.5,0.3)$ for the bottom line (a); $(0.76,0.2)$ for the middle line (b); and $(0.84,0.12)$ for the top line (c).
Four styles of the curves denote the four cases of $P_{\rm suc, max}$, which are i): $ \gamma^c_{1} \geq 1 $ (dotted), ii):  $ 1  \geq \gamma^c_{1}  \geq \gamma$ (dashed), iii): $ \gamma^c_{2}  \geq \gamma  \geq \gamma^c_{1}$ (solid), and iv): $ \gamma   \geq \gamma^c_{2} $ (dotted-dashed).
The inset shows $\Delta P=P_{\rm suc,max}-(1-\gamma)$.
} \label{fig2}
\end{figure}

Relations between optimal probabilities in (\ref{Pm1}-\ref{Pm4}) and the \emph{priori} overlap $\gamma$ for fixed values of $\{p_j \}$ in the three regions are plotted in Fig. \ref{fig2}.
Both the maximal success probabilities in regions (b) and (c) are larger than the one in region (a), and the advantages of the two cases are shown in the inset of Fig. \ref{fig2}.
This result can be explained by the fact that the \emph{priori} knowledge increases the optimal success probability.
The region (a) is the nearest one to the point $A$ with equal a \emph{priori} probabilities which corresponds to the least \emph{priori} knowledge for a fixed value of $\gamma$.
In this region, the optimal discrimination occurs when the states $|\Phi_j\rangle$ in (\ref{Utrans}) are the same.
The \emph{priori} knowledge increases as the probabilities $\{p_j \}$ depart from point $A$.
When the triangle relation $\gamma^c_1 > 1$ is violated, the states $|\Phi_j\rangle$ in the optimal discrimination are no longer the same.
And then $J_0<1< J_1 ,J_2 $, where we set $J_i = |J_{jk}/J_{ij}J_{k i}|$ and $(i,j,k)=(0,1,2),(1,2,0)$ or $(2,0,1)$.
That is, the states $|\psi_j\rangle$ are treated differently in the optimal discrimination according to their probabilities,
 even the states with low probabilities are omitted.



\section{Roles of correlations}
We are now ready to investigate the roles of correlations in the above AOSD by adjusting $|\eta_i\rangle$ without affecting the optimal success probability $P_{\rm suc, max}$.
We first show that entanglement can be absent in $\rho_{SA}$.
In the AOSD of two states, Zhang \emph{et. al.} \cite{Zhang} give a decomposition of the system-ancilla state with two separable pure states.
This inspires us to assume
\begin{align}\label{change}
\rho_{SA}=|\Psi_0\rangle\langle\Psi_0|+|\Psi_1\rangle\langle\Psi_1|+|\Psi_2\rangle\langle\Psi_2|,
\end{align}
and $|\Psi_{0,1,2}\rangle$ to be three separable pure states.
Let the pure states $|\Psi_i\rangle=\sum_{j=0}^2 c_{ij} \sqrt{p_j} \mathcal{U}\;|\psi_j\rangle|k\rangle_a$ with $i=0,1,2$, and $c_{ij}$ are the elements of an orthogonal matrix $C$ \cite{Wootters98}.
They can be written as
\begin{align}
|\Psi_i\rangle=|\mu_i\rangle|0\rangle+|\nu_i\rangle|1\rangle,
\end{align}
where $|\mu_i\rangle =\sum_{j=0}^2 c_{ij} \sqrt{p_j}  \sqrt{1-|\alpha_j|^2}|j\rangle $ and $|\nu_i\rangle = \sum_{j=0}^2 c_{ij} \sqrt{p_j}   \alpha_j|\Phi_j\rangle.$
Then, the problem to prove the absence of entanglement becomes to find the basis $\{ |\eta_i\rangle\}$ and the orthogonal matrix $C$ making $|\Psi_{0,1,2}\rangle$ separable.
Actually, only $|\eta_0\rangle$ and  $|\eta_1\rangle $ in the basis influence the state $\rho_{SA}$ and $\alpha_j$ can always be real in the optimal discrimination as show in the above section.

Let us first consider the cases with the optimal discrimination probabilities $P^{(ii)}_{\rm suc, max}$, in which the parameters $\theta_3=\pi$ and $\phi=0$.
The orthogonal matrix $C$ can be parameterize as $C=\Omega  \Lambda$ with
\begin{equation}
\Lambda =\left(
\begin{array}
{ccc}
\cos\kappa_1 & \sin\kappa_1\cos\kappa_2 & \sin\kappa_1\sin\kappa_2 \\
0 & -\sin\kappa_2  & \cos\kappa_2  \\
-\sin\kappa_1 & \cos\kappa_1\cos\kappa_2  & \cos\kappa_1\sin\kappa_2
\end{array}%
\right),
\end{equation}
and
\begin{equation}
\Omega =\left(
\begin{array}
{ccc}
1 & 0 & 0 \\
0 & \cos \kappa_3  & -\sin \kappa_3  \\
0 & \sin \kappa_3  & \cos \kappa_3
\end{array}%
\right).
\end{equation}
Since only two of the states $|\Phi_i\rangle$ are linearly independent, we assume $|\nu_0\rangle=0$, which leads to
\begin{eqnarray}
c_{00}:c_{01}:c_{02}=-\frac{\sin(\theta_1+\theta_2)}{\sqrt{p_0}\alpha_0}:\frac{\sin \theta_1 }{\sqrt{p_1}\alpha_1}:\frac{\sin \theta_2 }{\sqrt{p_2}\alpha_2}.
\end{eqnarray}
The parameters $\kappa_1$ and $\kappa_2$ can be determined by this relation.
Set $|\mu'_i\rangle =\sum_{j=0}^2 \lambda_{ij} \sqrt{p_j}  \sqrt{1-|\alpha_j|^2}|j\rangle $ and $|\nu'_i\rangle = \sum_{j=0}^2 \lambda_{ij} \sqrt{p_j}   \alpha_j|\Phi_j\rangle$ with $\lambda_{ij} $ being the elements of $\Lambda$,
 and define two orthonormalized states
\begin{eqnarray}\label{tau}
|\tau_0\rangle= \frac{|\mu'_1\rangle}{\sqrt{\langle \mu'_1|\mu'_1\rangle}}, \ \ \
|\tau_1\rangle=\frac{|\mu'_2\rangle- \langle \tau_0|\mu'_2\rangle|\tau_0\rangle}{\sqrt{\langle \mu'_2|\mu'_2\rangle - | \langle \tau_0|\mu'_2\rangle|^2}}.
\end{eqnarray}
A necessary condition for the separability of $|\Psi_{ 1 }\rangle$ and $|\Psi_{2}\rangle$ is that the states $|\eta_0\rangle$ and $|\eta_1\rangle$ are in the space expanded by  $|\tau_0\rangle$ and $|\tau_1\rangle$.
 For simplicity, we assume
 \begin{eqnarray}\label{beta}
|\eta_0\rangle=\cos\beta |\tau_0\rangle+\sin\beta|\tau_1\rangle,\
|\eta_1\rangle=-\sin\beta |\tau_0\rangle+\cos\beta|\tau_1\rangle.
\end{eqnarray}
Substituting them into $|\nu_1\rangle$ and $|\nu_2\rangle$, and requiring the $|\Psi_{ 1 }\rangle$ and $|\Psi_{2}\rangle$ to be separable, one has two equations of the parameter $\kappa_3$ as
\begin{eqnarray}\label{twoeq}
&&\frac{\cos \kappa_3 \mathcal{P} - \sin \kappa_3 \mathcal{Q}}{\cos \kappa_3 \mathcal{A} - \sin \kappa_3 \mathcal{B}_0} =\frac{\cos \kappa_3 \mathcal{R} - \sin \kappa_3 \mathcal{S}}{ - \sin \kappa_3 \mathcal{B}_1}, \nonumber\\
&&\frac{\sin\kappa_3 \mathcal{P} + \cos \kappa_3 \mathcal{Q}}{\sin \kappa_3 \mathcal{A} + \cos \kappa_3 \mathcal{B}_0} =\frac{\sin \kappa_3 \mathcal{R} + \cos \kappa_3 \mathcal{S}}{  \cos \kappa_3 \mathcal{B}_1},
\end{eqnarray}
where $\mathcal{A}=\langle \tau_0|\mu'_1\rangle$, $\mathcal{B}_0=\langle \tau_0|\mu'_2\rangle$, $\mathcal{B}_1=\langle \tau_1|\mu'_2\rangle$, $\mathcal{P}=\langle \tau_0|\nu'_1\rangle$, $\mathcal{Q}=\langle \tau_0|\nu'_2\rangle$, $\mathcal{R}=\langle \tau_1|\nu'_1\rangle$, and $\mathcal{S}=\langle \tau_1|\nu'_2\rangle$.
The compatibility condition of the two equations in (\ref{twoeq}) is
 \begin{eqnarray} \label{cond}
 \mathcal{AR}=\mathcal{QB}_1-\mathcal{SB}_0
 \end{eqnarray}
which is a linear equation of $\tan \beta$ and can be solved directly.
With this, we have shown that the state $\rho_{SA}$ with the optimal discrimination probabilities $P^{(ii)}_{\rm suc, max}$ is separable when the states $|\eta_0\rangle$ and $|\eta_1\rangle$ satisfy (\ref{beta}) and (\ref{cond}).

For the other cases, the states $|\eta_0\rangle$ and $|\eta_1\rangle$ corresponding to a separable  $\rho_{SA}$  can be derived in a similar way.
One also can obtain these results by calculating the limits of (\ref{beta}) and (\ref{cond}).
In the limit of $\theta_0 = \theta_1 \rightarrow 0$,
\begin{eqnarray}\label{cond1}
 |\eta_0\rangle \propto \sum^2_{i=0} p_i \alpha_i \sqrt{1-|\alpha_i|^2}|i \rangle.
\end{eqnarray}
This is the solution for the AOSD with $P^{(i)}_{\rm suc, max}$, where $|\eta_1\rangle$ disappears from $\rho_{SA}$. For the case with $\alpha_2=1$, the states in (\ref{tau}) becomes $|\tau_0\rangle=|1\rangle$ and $|\tau_1\rangle=|0\rangle$, and the condition (\ref{cond}) reduces to
\begin{eqnarray}\label{cond2}
p_0 \alpha_0 \sqrt{1-|\alpha_0^2|} \cos\beta = p_1 \alpha_1 \sqrt{1-|\alpha_1^2|}\sin{(\beta+\theta_1)}.\ \ \
\end{eqnarray}
When $\theta_1=0$, it returns to the result of the AOSD for the two states in \cite{Zhang}.
And, when $\alpha_1=1$, it leads to the solution for the case with $P^{(iv)}_{\rm suc, max}$ as
\begin{eqnarray}\label{cond3}
 |\eta_0\rangle=|0\rangle, \ \ |\eta_1\rangle=-|1\rangle .
  \end{eqnarray}

The above results reveal that AOSD of the three states with positive real overlaps can be performed in the absence of entanglement.
The recent developments \cite{roa2011dissonance,BoLi,Zhang} bring us to consider the dissonance as the key resource in this quantum information processing.
The quantum dissonance is equal to quantum discord for a separable state \cite{modi2010unified}.
However, the quantum discord does not have an analytic or operational expression in general, because of the supreme in its original definition  \cite{PhysRevLett.88.017901,henderson2001classical}.
In this work, we adopt the definition of GMQD \cite{GMQD,GMQD2013} as its amount, and calculate it for the case with equal a \emph{priori} overlaps.
For our bipartite system in the space $H^{S}\otimes H^{A}$ with dim $H^{S}=3$ and dim $H^{A}=2$,
the state (\ref{rho}) can be represented as
\begin{eqnarray}
\rho_{SA}&=&\frac{1}{6}\biggr(\openone_3 \otimes \openone_2 +\sum\limits_{i=1}^{3}x_{i} \openone_3\otimes \sigma_{i}+\sum\limits_{j=1}^{8}y_{j}\lambda_j\otimes \openone_2 \nonumber\\
& &\ \ \  \ \ \  +\sum\limits_{j=1}^{8} \sum\limits_{i=1}^{3}t_{ji}\lambda_j\otimes \sigma_{i}\biggr),
\end{eqnarray}
where $\openone_3$ and $\openone_2$ are the unit matrixes for respective dimension,
$\sigma_i$ and $\lambda_j$ are the traceless Hermitian generators of SU(2) and SU(3) respectively,
which satisfy ${\rm Tr}(\sigma_i\sigma_j)={\rm Tr}(\lambda_i \lambda_j)=2\delta_{ij}$.
The vectors $x_i$, $y_j$ and tensor $t_{ji}$ can be calculated as
\begin{eqnarray}
&&x_i= {\rm Tr}[\rho_{SA}(\openone_3\otimes \sigma_{i})], \ \ \ \
y_j={\rm Tr}[\rho_{SA}(\lambda_j\otimes \openone_2)], \ \ \   \ \ \  \nonumber\\
&&t_{ji}={\rm Tr}[\rho_{SA}(\lambda_j\otimes \sigma_{i})].
\end{eqnarray}
The GMQD \cite{GMQD,GMQD2013} can be given by
\begin{eqnarray}\label{GMQD}
\mathcal{D}_G(\rho_{SA})=\frac{1}{6}||X||^2+\frac{1}{4}||\tau||^2-k_{max},
\end{eqnarray}
where $X=(x_1,x_2,x_3)^{\rm T}$, $\tau$ is the matrix with elements $t_{ji}$,
and $k_{max}$ is the maximal eigenvalue of the matrix ($\frac{1}{6}XX^{\rm T}+\frac{1}{4}\tau \tau^{\rm T}$).

\begin{figure}
\includegraphics[width=8cm]{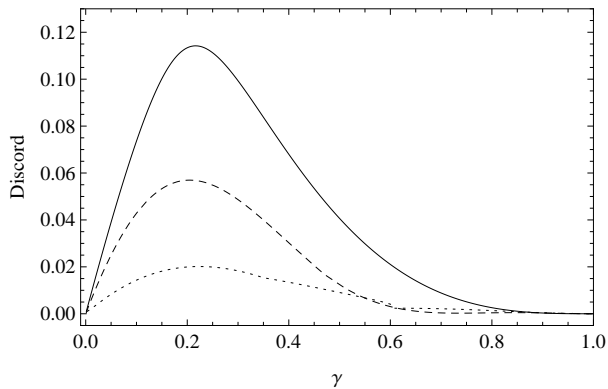} \\
 \caption{We plot the  quantity $2 \mathcal{D}_G(\rho_{SA})$ as functions of $\gamma$.
 The \emph{priori} probabilities are the same as the three lines in Fig. \ref{fig2}, chosen in the three regions in Fig. \ref{fig1}, in which the value of $(p_0,p_1)$ are (a): $(0.5,0.3)$ for solid line, (b): $(0.76,0.2)$ for dashed line, and (c): $(0.84,0.12)$ for dotted line.
} \label{fig3}
\end{figure}

Since the solution of $|\eta_i\rangle$ for the absence of entanglement in state $\rho_{SA}$ is not unique, we choose the ones satisfying the equations (\ref{cond}-\ref{cond3}),
 which provides a uniform treatment to the four cases of \emph{priori} probabilities and overlap.
Since the GMQD is not normalized to one and its value for maximally entangled qutrit-qubit states is $0.5$,
we plot the quantity $2 \mathcal{D}_G(\rho_{SA})$ for the case with equal a \emph{priori} overlaps in Fig. 3.
 The \emph{priori} probabilities are the same as the three lines in Fig. \ref{fig2}.
It is shown that the GMQD has a nonzero value for general case.
That is, the dissonance is a key ingredient in the quantum information processing of AOSD for the three states.
In addition, the amount of dissonance for the solid line corresponding to the region (a) in Fig. \ref{fig1} is larger than the other two cases.
The region (a) is more close to the point $A$, which corresponds to the maximal \emph{priori} entropy of the system qutrit.
This indicates the lack of \emph{priori} knowledge increases the requirement of quantum correlations. The same result has also been obtained in the AOSD of two states \cite{Zhang}.
To give a more explicit picture, let us consider the discrimination with the same $|\Phi_i\rangle=|\eta_0\rangle$, which has the success probability $P_{\rm suc}=1-\gamma$.
This is the optimal protocol in region (a) and close to the ones for the other two regions as shown in Fig. \ref{fig2}.
According to our scheme to show the absence of entanglement above, one can write the separable system-ancilla state as
\begin{eqnarray}
\rho_{SA}= \rho_1 \otimes q |0\rangle_a \langle0|+ |\eta_0\rangle \langle\eta_0|\otimes |\xi\rangle_a \langle\xi|,
\end{eqnarray}
where $\rho_1 $ is a  normalized state of the qutrit, $q=2(1-\gamma)(p_0 p_1 + p_1 p_2+ p_2 p_0)$, $|\eta_0\rangle=\sum_{j=0}^2 p_j|j\rangle/\sqrt{\sum_{i=0}^2 p^2_i}$, and $ |\xi\rangle_a=\sqrt{1-\gamma}\sqrt{\sum_{i=0}^2 p^2_i}|0\rangle_a+\sqrt{\gamma}|1\rangle_a$.
The necessary and sufficient condition of zero discord in $\rho_{SA}$ is the commutator
\begin{eqnarray}
&&[ q |0\rangle_a \langle0|,|\xi\rangle_a \langle\xi|]  \nonumber \\
&&=q\sqrt{\gamma(1-\gamma) (p_0^2 +p_1^2 +p_2^2)} (|0\rangle_a \langle1|-|1\rangle_a \langle0|),\ \ \ \ \ \
\end{eqnarray}
to be zero \cite{GMQD}.
For a fixed value of $\gamma$, the coefficient in the commutator reaches its maximum when \emph{priori} probabilities are the same,
 which can be considered as a symbol of the maximal discord.

\section{Summary}
We have studied the protocol for unambiguous discrimination of three nonorthogonal states assisted by a qubit and explored the roles of quantum correlations in this quantum information procedure.
We confined ourselves to the case with positive real overlaps, which is simplest example to extend the results in \cite{roa2011dissonance,Zhang} to a high dimensional system.
Although the analysis of optimal discrimination is more complicated than the case of two states, the entanglement is proved to be completely unnecessary in the optimal procedure.
We also calculated the dissonance in the optimal discrimination of the case with equal a \emph{priori} overlaps, by using the definition of geometric discord.
It was shown that the \emph{priori} knowledge deduces the requirement of dissonance.
These results reveal that the original results in \cite{roa2011dissonance,Zhang} about the roles of entanglement and dissonance in AOSD may come into existence in arbitrary dimension with  arbitrary  \emph{priori} probabilities.

Unambiguous discrimination among linearly independent nonorthogonal quantum states is important in both of quantum mechanics and quantum information theory.
It is employed in, e.g., the protocol of \emph{conclusive quantum teleportation} \cite{Zhang,kim2004generalized} where the resource is not prepared in a maximally entangled state.
Very recently, based on unambiguous discrimination, the topic of extracting information from a qubit by multiple observers has been studied \cite{bergou2013extracting,pang2013sequential}.
Most of the existing results about optimal unambiguous discrimination were derived in the POVM formalism \cite{PhysRevA.82.032338,POVM2008,POVM2009}, which can be realized by introducing
ancillary systems and performing von Neumann measurements on systems and the ancillas.
The present work and previous studies \cite{roa2011dissonance,BoLi,Zhang,pang2013sequential} investigating unambiguous quantum state discrimination by using the language of system-ancilla
provides a way to understand the roles of quantum correlations in the quantum information process.

\acknowledgments
F.L.Z. is supported by NSF of China (Grant No. 11105097). J.L.C. is
supported by National Basic Research Program (973 Program) of China
under Grant No. 2012CB921900, NSF of China (Grant Nos. 10975075 and
11175089) and partly supported by National Research Foundation and
Ministry of Education of Singapore.

\bibliography{AOSD_EPL}

\end{document}